\documentclass[aps,pre,groupedaddress,showpacs,preprint]{revtex4}
\usepackage{graphicx}% Include figure files
\usepackage[dvips]{epsfig}% Include figure files
\usepackage{dcolumn}% Align table columns on decimal point
\usepackage{subfigure}%
\usepackage{bm}% bold math
\usepackage{amssymb}
\usepackage{amsmath}

\begin{document}
\title{Persistence of instanton connections in chemical reactions with time dependent rates}

\author{Carlos Escudero$^\dag$ and Jos\'{e} \'{A}ngel Rodr\'{\i}guez$^\ddag$}

\affiliation{$\dag$ Mathematical Institute, University of Oxford, 24-29 St Giles', Oxford OX1 3LB, United Kingdom \\
$\ddag$ Departamento de Matem\'{a}ticas, Universidad de Oviedo, Avenida de Calvo Sotelo s/n, 33007 Oviedo, Spain}

\begin{abstract}
The evolution of a system of chemical reactions can be studied, in the eikonal approximation, by means of a Hamiltonian dynamical system. The fixed points of this dynamical system represent the different states in which the chemical system can be found, and the connections among them represent instantons or optimal paths linking these states. We study the relation between the phase portrait of the Hamiltonian system representing a set of chemical reactions with constant rates and the corresponding system when these rates vary in time. We show that the topology of the phase space is robust for small time-dependent perturbations in concrete examples and state general results when possible. This robustness allows us to apply some of the conclusions on the qualitative behavior of the autonomous system to the time-dependent situation.
\end{abstract}

\pacs{05.40.-a, 05.45.-a, 64.60.My, 82.20.-w}
\maketitle

\section{Introduction}

Understanding the dynamics of chemical kinetics is a question of fundamental character in nonequilibrium statistical mechanics, apart from being of broad importance in applications to other sciences. Indeed, many models in chemistry~\cite{mastny}, biochemistry~\cite{kepler}, ecology~\cite{escudero1}, and biology~\cite{condat} use stoichiometric relations as theoretical first principles describing some phenomenon. The simplest description of the time evolution of a set of $N$ reacting species is probably given by the mean-field equations, an $N-$dimensional dynamical system representing the concentrations, or total number of molecules of the reacting species. One of the advantages of this approach is that it allows us to use the powerful machinery of dynamical systems theory~\cite{holmes}, raising the possibility of identifying stationary states with fixed points, periodic behavior with limit cycles, etc.

Of course, as with every theory in physics, the mean-field approximation has a range of validity. As it ignores fluctuations, its description of the chemical system might be accurate for short times; however, long-time dynamics will be affected, dramatically in some cases, by rare events. It is possible to study exactly the system evolution as a time-continuous Markov process, the probability distribution of which is given by the solution of an adequate master equation~\cite{gardiner,kampen}. The full analytical solution of the master equation is, in most situations, a formidable problem, and it yields, usually, much more information than that needed in applications. The development of approximate theories is thus clearly justified.

One of the most common approximations is the Fokker-Planck equation, obtained from the master equation by means of a Kramers-Moyal or van Kampen size expansion~\cite{gardiner,kampen}. This theory assumes that the number of reagents is very large, so we can consider the implicit stochasticity of the process as small Gaussian fluctuations around the mean-field behavior. Definitely, rare events do not belong to this category. If one wants to deal with fluctuations that are comparable with, or even greater than, the mean value, any approximation scheme must not rely on assumptions such as an asymptotically large value of the number of reagents. It is possible to construct such a theory, for instance, by taking advantage of the relation among chemical kinetics and quantum mechanics~\cite{doi,peliti}. Once the master equation is formulated as a quantum problem, it is possible to develop eikonal~\cite{dykman1,elgart} or WKB approximations~\cite{assaf}, or matched asymptotic expansions of the spectral formulation~\cite{assaf2}, able to tackle rare events. In this work we will concentrate on the eikonal approximation introduced in~\cite{elgart}. One of the advantages of this approach is that it reduces the problem again to a dynamical system of $2N$ dimensions for $N$ reacting species. The additional $N$ degrees of freedom are the conjugate "momenta" corresponding to each of the concentrations and represent a measure of the size of the fluctuations. Due to the Hamiltonian symmetry of this system, it can be effectively reduced to a $(2N-1)-$dynamical system on some Riemannian manifold. So, if there is only one chemical species, the case in which we will concentrate here, we have a reduced number of dynamical scenarios. In particular, only fixed points will be of physical relevance, and they will denote the possible stationary states in which the system can be found. Contrary to the mean-field situation, rare events can drive the system for one stationary state to another, a fact that is reflected by the existence of connections between the fixed points of the dynamical system. These connections are not the unique, but the optimal way in which a system evolves from one state to the other~\cite{dykman1}, and, as they are reminiscent of instantons in quantum mechanics~\cite{zinn}, we will denote them as instanton connections. Its importance is huge: the web of connections, having the fixed points at their intersections, encodes the qualitative behavior of the chemical system~\cite{elgart}, and its topology serves as a principle for the classification of nonequilibrium phase transitions in reaction-diffusion models~\cite{elgart2}.

All these works have been focused on the dynamics of chemical kinetics happening at constant rates. While this assumption is reasonable in many cases, there are situations in which we should go beyond it and consider the explicit time variation of the reaction rates. Periodically illuminated chemical reactions~\cite{epstein} or seasonal variation in population dynamics~\cite{buceta} serve as examples of nontrivial behavior generated by a temporal forcing. Not much attention seems to have been paid to eikonal approximations of time-dependent chemical kinetics, and when they are considered, nonautonomous perturbations are usually treated within the quasistationary approximation~\cite{dykman2,escudero}. It is our goal to extend the existent approaches and consider arbitrary frequency, albeit small, perturbations. The concrete problem under study is how the phase portrait of the eikonal dynamical system is modified when time-dependent perturbations enter into play: do the instanton connections survive or do they disappear changing the qualitative behavior of the system? Giving a totally general answer to this question is a difficult task, but we will see that these connections seem to be very robust and persistent when they are subject to a small periodic forcing. This will be shown with the help of particular models, for which rigorous results (for the eikonal approximation) are easily proven, and then we will extend them to the general setting when possible.

\section{Instanton Persistence}
\label{ip}

\begin{figure}
\begin{center}
\psfig{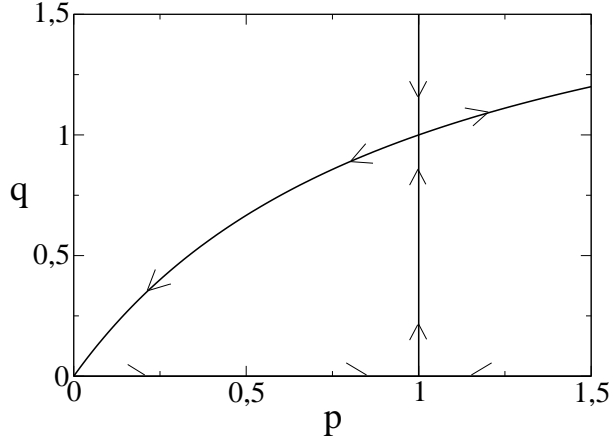}
\end{center}
\caption{Phase space for the branching and annihilating particle problem. The parameter values are $\sigma=\lambda=1$.}
\label{fase1}
\end{figure}

\subsection{Branching and annihilation}

For the shake of clarity, we will illustrate the problem of instanton persistence with a particular reaction, branching and annihilation of identical particles, but we will state general results for arbitrary reaction sets~\cite{mc}. Let us start considering a single species of identical particles $A$, which annihilate in pairs and undergo binary branching
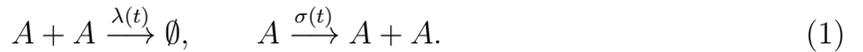
\begin{equation}
A+A \stackrel{\lambda(t)}{\longrightarrow} \emptyset, \qquad A \stackrel{\sigma(t)}{\longrightarrow} A+A.
\end{equation}
The master equation describing the probability distribution of having $n$ reagents at time $t$ reads
\begin{equation}
\frac{dP_n(t)}{dt}=\sigma(t)[(n-1)P_{n-1}(t)-nP_n(t)]+\frac{\lambda(t)}{2} [(n+2)(n+1)P_{n+2}(t)- n(n-1)P_n(t)].
\end{equation}
It can be represented by a partial differential equation (PDE) by introducing the generating function
\begin{equation}
G(p,t) \equiv \sum_{n=0}^\infty p^n P_n(t),
\end{equation}
which provides us with the time-dependent Hamiltonian
\begin{equation}
H(p,q,t)=\sigma(t)(p-1)pq+\frac{\lambda(t)}{2}(1-p^2)q^2
\end{equation}
and the imaginary time Schr\"{o}dinger equation
\begin{equation}
\partial_t G = -H(p,-\partial_p,t)G.
\end{equation}
The eikonal approximation proposes the reduction of the problem to Hamilton equations~\cite{elgart}, which in this case become the nonautonomous dynamical system (note, however, that we could reduce the problem to one time-dependent reaction rate by means of the time reparametrization $\tilde{t}=\int_0^t \sigma (\tau) d\tau$)
\begin{eqnarray}
\label{brann1}
\dot{p} &=& -\frac{\partial H}{\partial q} =\sigma(t)(1-p)p +\lambda(t)(p^2-1)q, \\
\dot{q} &=& \frac{\partial H}{\partial p} =\sigma(t)(2p-1)q - \lambda(t)pq^2.
\label{brann2}
\end{eqnarray}
Suppose for a moment that both $\sigma$ and $\lambda$ are time independent. In this case the system exhibits three lines with zero energy: the invariant lines $p=1$, $q=0$, and
\begin{equation}
q=\frac{2\sigma p/ \lambda}{1+p}.
\end{equation}
Furthermore, we know that these three lines connect the fixed points $(0,0)$, $(1,0)$, and $(1,\sigma/\lambda)$ in the $(p,q)$ plane, see Fig.(\ref{fase1}) (or Fig.(3) in~\cite{elgart}). Now we can try to understand what happens when we let the reaction rates vary in time. It is easy to see that the lines $p=1$ and $q=0$ remain invariant zero-energy lines of the system; however, the explicit time dependence of the Hamiltonian prevents the conservation of the energy $H$ and, in general, forces the disappearance of the third zero-energy line. We can generalize this fact for an arbitrary reaction Hamiltonian. Given any set of reactions with time-dependent rates, we know that the Hamiltonian necessarily fulfills
\begin{equation}
H(p=1,q,t)=0
\end{equation}
due to the conservation of probability~\cite{elgart,elgart2}. So this means that \emph{for any set of (time-dependent) reaction rules, the $p=1$ line is an invariant, zero-energy line of the dynamical system}. Since this line describes the mean-field dynamics of the system, we will call it the \emph{mean-field line}~\cite{elgart,escudero}. Some systems possess an absorbing state when they contain zero particles; this happens if all reactions of the type
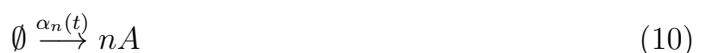
\begin{equation}
\emptyset \stackrel{\alpha_n(t)}{\longrightarrow} n A
\end{equation}
are absent in the dynamics. In this case, the Hamiltonian must obey the condition~\cite{elgart,elgart2}
\begin{equation}
H(p,q=0,t)=0.
\end{equation}
So we can claim that \emph{any (nonautonomous) system without particle production from the vacuum has the invariant, zero-energy line $q=0$}. And thus, when it is present, we will call it \emph{the absorbing-state line}. These properties can be clearly seen from the general form of the Hamiltonian term representing the reaction
\begin{equation}
mA \stackrel{\beta_{mn}(t)}{\longrightarrow} nA;
\end{equation}
it is~\cite{elgart2}
\begin{equation}
\label{genreact}
H_{mn} = \frac{\beta_{mn}(t)}{m!}(p^n-p^m)q^m.
\end{equation}
In the autonomous situation, the set of zero-energy lines determines the physics of the problem: the fixed points obtained when these lines cross represent the possible states in which the system can be found and the connections among them the possible transition paths. Together with the mean-field line (the global minimum of the action) and the absorbing-state line one finds other lines with zero energy: \emph{the instanton lines}~\cite{elgart,elgart2,escudero}. We know that both the
mean-field line and the absorbing-state line persist if we let the reaction rates vary in time, but however, it is not so obvious to see what happens with the instanton lines. These are defined in terms of zero energy if the system is not explicitly dependent on time, but when this is not the case, the definition loses its meaning since energy is no longer conserved. Due to this fact and because the physical role of the instanton lines is to be optimal paths between different states, what we would like to know at this point is if both fixed points and connections among them survive after the nonautonomous forcing is switched on. We can be sure that hyperbolic fixed points persist to a small periodic time-dependent forcing, after possible relocation of their position, but do the instanton connections persist?

\begin{figure}
\begin{center}
\psfig{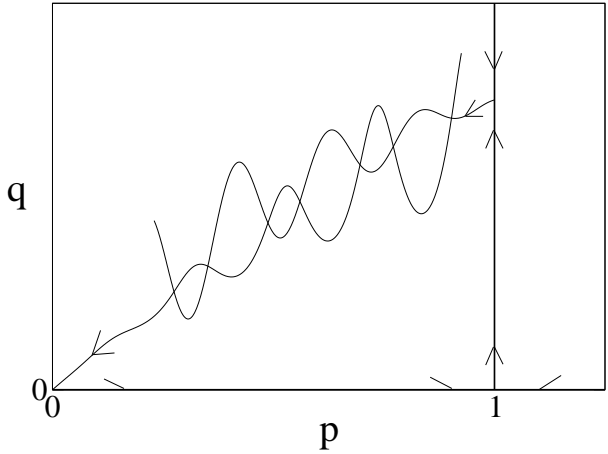}
\end{center}
\caption{Sketch of the phase space for the branching and annihilating particle problem, assuming that the stable and unstable branches intersect. The heteroclinic connection goes through all the intersections of these branches.}
\label{fase2}
\end{figure}

\begin{figure}
\begin{center}
\psfig{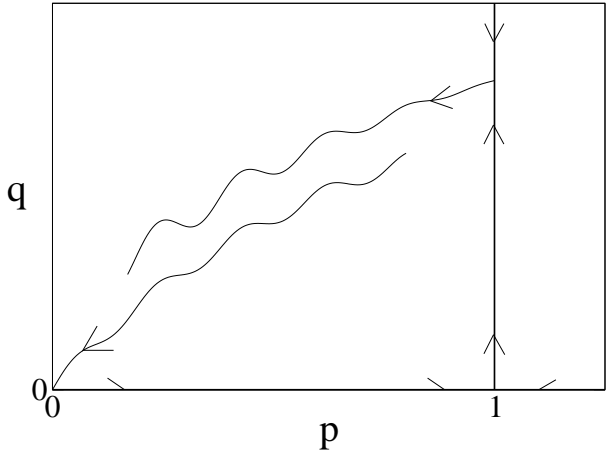}
\end{center}
\caption{Sketch of the phase space for the branching and annihilating particle problem, assuming that the stable and unstable branches do not intersect.}
\label{fase3}
\end{figure}

To address this question we will use the method developed by Melnikov~\cite{melnikov} (see also~\cite{holmes}). Consider an autonomous two-dimensional dynamical system
\begin{equation}
\dot{{\bf x}}={\bf f}({\bf x}),
\end{equation}
${\bf x}=(x_1,x_2)$, ${\bf f}=(f_1,f_2)$, with two hyperbolic fixed points ${\bf x}_a$ and ${\bf x}_b$ linked by a heteroclinic connection, which is parametrized by the system solution ${\bf x}_h(t-t_0)$ for initial time $t_0$. Assuming that the motion goes from ${\bf x}_a$ to ${\bf x}_b$, this connection is formed by a branch of the unstable manifold of ${\bf x}_a$, which totally overlaps with a branch of the stable manifold of ${\bf x}_b$. Let us now consider the perturbed version of this problem,
\begin{equation}
\label{perturbed}
\dot{{\bf x}}={\bf f}({\bf x})+\epsilon {\bf g}({\bf x},t),
\end{equation}
where the perturbation ${\bf g}({\bf x},t)$ is time periodic with period $T$, amplitude $\epsilon$ small enough and sufficiently regular. The dynamics of this nonautonomous system is given by the associated Poincar\'{e} map $\mathcal{P}_\epsilon$, which maps every initial condition point ${\bf x}(0)$ with the corresponding value
of the solution ${\bf x}(T)$ after one period has elapsed~\cite{holmes}. The hyperbolic fixed points of the unperturbed system, ${\bf x}_a$ and ${\bf x}_b$, are hyperbolic fixed points of the Poincar\'{e} map $\mathcal{P}_0$. Since the map $\mathcal{P}_\epsilon$ is a perturbation of $\mathcal{P}_0$, these points have a continuation, ${\bf x}_a^\epsilon$ and ${\bf x}_b^\epsilon$, as hyperbolic fixed points of $\mathcal{P}_\epsilon$, and their invariant manifolds vary continuously with respect to $\epsilon$. The perturbed system will not, in general, maintain the coincidence between the branches of the unstable and stable manifolds of ${\bf x}_a$ and
${\bf x}_b$, respectively: now these branches might intersect, preserving the existence of the heteroclinic connection (see Fig.(\ref{fase2})) or might not, destroying it, like in Figs.(\ref{fase3}) and (\ref{fase4}). The distance between the stable and unstable branches is given by $d(\epsilon,t_0)=\epsilon M(t_0)+O(\epsilon^2)$, with
\begin{equation}
\label{melnikov}
M(t_0)=\int_{-\infty}^{\infty} {\bf f}({\bf x}_h(t-t_0)) \wedge {\bf g}({\bf x}_h(t-t_0),t)dt,
\end{equation}
where
\begin{equation}
{\bf f} \wedge {\bf g} = \left| \begin{array}{cc}
f_1 & f_2 \\
g_1 & g_2 \end{array} \right|
\end{equation}
denotes the wedge product of vectors ${\bf f}$ and ${\bf g}$. Equation~(\ref{melnikov}) defines the so-called Melnikov function, which yields the first-order approximation in $\epsilon$ of the distance between the stable and unstable manifolds measured along a direction that is perpendicular to the unperturbed connection at the point ${\bf x}_h(t_0)$. A change of sign of $M(t_0)$ means that there exists some $t_0$ such that $d(\epsilon,t_0)=0$, implying the existence of a solution ${\bf x}_h^\epsilon(t)$ of Eq.~(\ref{perturbed}) defining a heteroclinic connection among the two hyperbolic fixed points ${\bf x}_a^\epsilon$ and ${\bf x}_b^\epsilon$ of the Poincar\'{e} map corresponding to Eq.~(\ref{perturbed}), say,
\begin{equation}
\lim_{t \to -\infty}{\bf x}_h^\epsilon(t)={\bf x}_a^\epsilon, \qquad \lim_{t \to \infty}{\bf x}_h^\epsilon(t)={\bf x}_b^\epsilon.
\end{equation}

Let us now consider the branching and annihilating system, Eqs.~(\ref{brann1}) and (\ref{brann2}), with constant reaction rates. The instanton connection (the separatrix linking $(1,\sigma/\lambda)$ to $(0,0)$) is parametrized by the system solution
\begin{equation}
{\bf x}_h(t-t_0)= (p_h(t-t_0),q_h(t-t_0)) = \left( \frac{1}{1+e^{\sigma(t-t_0)}},\frac{2\sigma/\lambda}{2+e^{\sigma(t-t_0)}} \right).
\end{equation}
If we assume that the perturbation of the reaction rates is given by
\begin{eqnarray}
\sigma(t) &=& \sigma + \epsilon \sigma_1(t), \\
\lambda(t) &=& \lambda + \epsilon \lambda_1(t),
\end{eqnarray}
and the new rates are $-$periodic and regular enough, we obtain the system
\begin{eqnarray}
\label{p1}
\dot{p} &=& \sigma(1-p)p +\lambda(p^2-1)q + \epsilon \left[ \sigma_1(t)(1-p)p +\lambda_1(t)(p^2-1)q \right], \\
\dot{q} &=& \sigma(2p-1)q - \lambda pq^2 + \epsilon \left[ \sigma_1(t)(2p-1)q - \lambda_1(t) pq^2 \right],
\label{q1}
\end{eqnarray}
which is of type~(\ref{perturbed}). The associated Poincar\'{e} map $\mathcal{P}_\epsilon$ still has the origin as hyperbolic fixed point. Another hyperbolic fixed point of this map, corresponding to $(1,\sigma/\lambda)$ when $\epsilon=0$, is $(1,q_\epsilon)$, where $q_\epsilon$ is obtained through the solution of Eq.~(\ref{q1}) for $p=1$. This is nothing but a Bernoulli differential equation, which can be straightforwardly integrated to get
\begin{equation}
q(t)=\frac{q(0)\exp \left[ \int_0^t \sigma(\tau)d\tau \right]}{1+q(0)\int_0^t \exp \left[\int_0^s \sigma(\tau)d\tau \right] \lambda(s) ds}.
\end{equation}
We just need to apply the condition $q(0)=q(T)$ to this solution to find
\begin{equation}
q_\epsilon= \frac{\exp \left[ \int_0^T \sigma(\tau)d\tau \right]-1}{\int_0^T \exp \left[ \int_0^s \sigma(\tau)d\tau \right]\lambda(s)ds}.
\end{equation}

\begin{figure}
\begin{center}
\psfig{file=fase4.eps,width=8cm,angle=0}
\end{center}
\caption{Sketch of the phase space for the branching and annihilating particle problem, assuming that the stable and unstable branches do not intersect.}
\label{fase4}
\end{figure}

To determine if the unstable manifold of $(1,q_\epsilon)$ intersects the stable manifold of $(0,0)$ we substitute the corresponding values of ${\bf f}$ and ${\bf g}$ into the Melnikov function~(\ref{melnikov}),
\begin{eqnarray}
\nonumber
M(t_0)=\int_{-\infty}^\infty \left[\lambda \sigma_1(t) - \sigma \lambda_1(t)\right]
\left[ (1-p)(1-p-p^2)q^2 \right]({\bf x}_h(t-t_0))dt= \\
\int_{-\infty}^\infty h(t)\phi(t-t_0)dt=\int_{-\infty}^\infty h(s/\sigma + t_0)\tilde{\phi}(s)ds,
\end{eqnarray}
after the change of variables $s=\sigma(t-t_0)$ and where
\begin{eqnarray}
h(t) &=& \lambda \sigma_1(t)-\sigma \lambda_1(t), \\
\phi(t) &=& \left[ (1-p)(1-p-p^2)q^2 \right]({\bf x}_h(t)), \\
\tilde{\phi}(s) &=& -\frac{4 \sigma^2}{\lambda^2}\frac{e^{2s}[1+2 \mathrm{sh}(s)]}{(1+e^s)^3(2+e^s)^2},
\end{eqnarray}
the symbol $\mathrm{sh}$ standing for the hyperbolic sine. It is easy to see that $\tilde{\phi}(s)$ has zero mean,
\begin{equation}
\int_{-\infty}^\infty \tilde{\phi}(s)ds=-\frac{4 \sigma^2}{\lambda^2}\int_{-\infty}^{\infty} \frac{e^{2s}[1+2 \mathrm{sh}(s)]}{(1+e^s)^3(2+e^s)^2} ds= \frac{2 \sigma^2}{\lambda^2}\left[ \frac{e^s}{(1+e^s)^2(2+e^s)}
\right]_{-\infty}^\infty=0,
\end{equation}
a fact that will prove its usefulness later on. Since $h(t)$ is $T$ periodic and continuous, we can expand it in Fourier series,
\begin{equation}
\label{phifourier}
h(t)=\sum_{n=-\infty}^{\infty} a_n e^{2 \pi i n t/T},
\end{equation}
and substitute it into the Melnikov function to obtain
\begin{equation}
\label{fourier}
M(t_0)=-\frac{4 \sigma^2}{\lambda^2} \sum_{n=-\infty}^{\infty} a_n e^{2 \pi i n t_0/T}
\int_{-\infty}^{\infty} \frac{e^{2s}[1+2 \mathrm{sh}(s)]}{(1+e^s)^3(2+e^s)^2}  e^{2 \pi i n s/(\sigma T)} ds,
\end{equation}
which is the Fourier decomposition of the Melnikov function. We can check the validity of the Fourier series~(\ref{fourier}) after contrasting that the norm
\begin{equation}
\int_{-\infty}^{\infty} \left| \frac{e^{2s}[1+2 \mathrm{sh}(s)]}{(1+e^s)^3(2+e^s)^2} \right| ds = \frac{5\sqrt{5}-11}{2}
\end{equation}
is finite. This representation allows us to calculate the mean value
\begin{equation}
\frac{1}{T}\int_{0}^T M(t_0)dt_0=a_0\int_{-\infty}^\infty \tilde{\phi}(s)ds=0,
\end{equation}
where we have used
\begin{equation}
\frac{1}{T}\int_{0}^T e^{2 \pi i n t_0/T}dt_0=\delta_{n0},
\end{equation}
and $\delta_{n0}$ denotes the Kronecker delta. So we know that the Melnikov function is periodic, continuous (from its very definition in Eq.~(\ref{melnikov}) it only depends on ${\bf f}$, ${\bf g}$, and ${\bf x}_h$), and with zero mean, implying that it either crosses zero or vanish identically. In the first case, we can claim that the point $(1,q_\epsilon)$ is connected to the origin, in the second, we cannot, due to the possible presence of small terms $O(\epsilon^2)$. However, we can see that the second case only happens when $h(t)$ is constant. Indeed, integrating by parts and changing variables $x=e^s$ we obtain
\begin{eqnarray}
\nonumber
\int_{-\infty}^{\infty} \frac{e^{2s}[1+2 \mathrm{sh}(s)]}{(1+e^s)^3(2+e^s)^2}  e^{2 \pi i n s/(\sigma T)} ds=
\frac{i \pi n}{\sigma T} \int_{-\infty}^{\infty} \frac{e^s}{(1+e^s)^2(2+e^s)}e^{2 \pi i n s/(\sigma T)}ds= \\
\frac{i \pi n}{\sigma T} \int_{0}^{\infty} \frac{x^{2 \pi i n/(\sigma T)}}{(1+x)^2(2+x)}dx.
\end{eqnarray}
This last integral can be obtained by means of contour integration in the complex plane. Noticing that for a complex variable $z$
\begin{equation}
z^{2 \pi i n/(\sigma T)}=e^{2 \pi i n \ln(z)/(\sigma T)},
\end{equation}
we can use the keyhole contour choosing the logarithm branch cut as the positive real axis, and by employing the residue theorem we obtain
\begin{equation}
\int_{0}^{\infty} \frac{x^{2 \pi i n/(\sigma T)}}{(1+x)^2(2+x)}dx=\frac{2 \pi i}{1-e^{-4 \pi^2 n/(\sigma T)}}
\{ \mathrm{Res}[f(z),-1]+\mathrm{Res}[f(z),-2] \},
\end{equation}
where
\begin{equation}
f(z)=\frac{z^{2 \pi i n/(\sigma T)}}{(1+z)^2(2+z)}.
\end{equation}
We can now compute the residues
\begin{eqnarray}
\mathrm{Res}[f(z),-2] &=& \exp \left[ -\frac{2 \pi^2 n}{\sigma T}+\frac{2 \pi i n}{\sigma T}\ln(2) \right], \\
\mathrm{Res}[f(z),-1] &=& \left. \frac{d}{dz}\frac{z^{2 \pi i n/(\sigma T)}}{(2+z)} \right|_{-1}=
-\left( 1+\frac{2 \pi i n}{\sigma T} \right) \exp \left( -\frac{2 \pi^2 n}{\sigma T} \right),
\end{eqnarray}
to conclude
\begin{eqnarray}
\nonumber
\int_{-\infty}^{\infty} \frac{e^{2s}[1+2 \mathrm{sh}(s)]}{(1+e^s)^3(2+e^s)^2}  e^{2 \pi i n s/(\sigma T)} ds= \\
\frac{2 \pi^2 n/(\sigma T)}{1-\exp[-4 \pi^2 n/(\sigma T)]}\exp \left( -\frac{2 \pi^2 n}{\sigma T} \right)
\left( 1+\frac{2 \pi i n}{\sigma T} -\exp \left[ \frac{2 \pi i n}{\sigma T}\ln(2) \right] \right),
\end{eqnarray}
and one can see that it only vanishes if $n=0$ or $n \to \pm \infty$, for $\sigma$ and $T$ positive finite real numbers. Hence, the Melnikov function is identically zero if and only if $a_n=0$ for all $n \neq 0$ in Eq.~(\ref{phifourier}) or, what is the same, if and only if $h(t)$ is constant.

We still have to analyze what is the meaning of the condition
\begin{equation}
\label{condition}
\lambda \sigma_1(t)-\sigma \lambda_1(t)=c,
\end{equation}
for some constant $c$. If $c=0$, then the system can be reduced to the unperturbed one by means of a reparametrization of the time variable:
\begin{equation}
u=\int_0^t \left[ 1+\frac{\sigma_1(\tau)}{\sigma} \right]d\tau.
\end{equation}
In this case the geometry of the phase space is exactly preserved; only the parametrization of the system solution along the separatrices might change. If $\sigma_1$ and
$\lambda_1$ are chosen constant, then for any value of $c$, the phase-space topology is the same, as a small retuning of the Hamiltonian parameters cannot modify it. In these two particular cases we know that the Melnikov function is identically zero because the connection is still given by the complete superposition of a branch of the stable manifold of one of the fixed points and a branch of the unstable manifold of the other. In the general case we know that a perturbation fulfilling $h(t)=c$ is a combination of these two, translation and reparameterization,
\begin{eqnarray}
\label{rep1}
\sigma(t) &=& (\sigma + \epsilon \sigma_1)[1+\epsilon \phi(t)], \\
\lambda(t) &=& (\lambda + \epsilon \lambda_1)[1+\epsilon \phi(t)],
\label{rep2}
\end{eqnarray}
for some function $\phi(t)$, up to terms $O(\epsilon^2)$. This means that, in order to know if the distance between the invariant manifolds of the fixed points is identically zero, we would have to use the extension of the Melnikov method to higher orders~\cite{battelli}.

\subsection{General setting}

It would be highly desirable to extend two properties of this example to general systems, say, the possibility of expressing the Melnikov function as a Fourier series, and to show that it has zero mean. We will start with the second claim assuming that the first one is true, and we will check its validity afterwards. The most general reaction Hamiltonian can be built adding the generic terms~(\ref{genreact}),
\begin{equation}
\mathcal{H} = \sum_{m \neq n} \frac{\beta_{mn}(t)}{m!}(p^n-p^m)q^m, \qquad m,n=0,1,2,\cdots,
\end{equation}
and it allows us to express the dynamical system as
\begin{eqnarray}
\label{perturbs1}
\dot{p} &=& \sum_{m \neq n} \frac{\beta_{mn}(t)}{(m-1)!}(p^m-p^n)q^{m-1}, \qquad m,n=0,1,2,\cdots, \\
\dot{q} &=& \sum_{m \neq n} \frac{\beta_{mn}(t)}{m!}(np^{n-1}-mp^{m-1})q^m, \qquad m,n=0,1,2,\cdots,
\label{perturbs2}
\end{eqnarray}
where all the rates $\beta_{mn}(t)$ are supposed to have the same period $T$ (the case of perturbations with different periods will be discussed below). Assuming the form of the perturbation $\beta_{mn}(t)=\beta_{mn}^0+ \epsilon \beta_{mn}^1(t)$ allows us to split the Hamiltonian into two terms $\mathcal{H} =\mathcal{H}_0 + \epsilon \mathcal{H}_1$. So we can write the integrand of the Melnikov function as the set of Poisson brackets
\begin{equation}
{\bf f}({\bf x}_h(t-t_0)) \wedge {\bf g}({\bf x}_h(t-t_0),t)=\{\mathcal{H}_1({\bf x}_h(t-t_0),t) , \mathcal{H}_0({\bf x}_h(t-t_0))\}
\end{equation}
and obtain
\begin{eqnarray}
\nonumber
M(t_0)=\int_{-\infty}^{\infty} \{\mathcal{H}_1({\bf x}_h(t-t_0),t) , \mathcal{H}_0({\bf x}_h(t-t_0)) \} dt= \\
\nonumber
\int_{-\infty}^\infty \frac{d}{dt}\mathcal{H}_1({\bf x}_h(t-t_0),t)dt-\int_{-\infty}^\infty \frac{\partial}{\partial t}\mathcal{H}_1({\bf x}_h(t-t_0),t)dt= \\
-\int_{-\infty}^\infty \frac{\partial}{\partial t}\mathcal{H}_1({\bf x}_h(t-t_0),t)dt,
\end{eqnarray}
where we have used the fact that instanton lines link fixed points lying on zero-energy invariant lines. Since
\begin{equation}
\mathcal{H}_1({\bf x}_h(t-t_0),t)= \sum_{m \neq n} \beta^1_{mn}(t) \mathcal{J}_{mn}(t-t_0), \qquad \mathcal{J}_{mn}(t-t_0)=\frac{1}{m!}(p(t-t_0)^n-p(t-t_0)^m)q(t-t_0)^m,
\end{equation}
we can integrate by parts to get
\begin{equation}
\label{mj}
M(t_0)= -\int_{-\infty}^\infty \sum_{m \neq n}\frac{d}{dt} \beta^1_{mn}(t) \mathcal{J}_{mn}(t-t_0) dt = \int_{-\infty}^\infty \sum_{m \neq n} \beta^1_{mn}(t) \frac{d}{dt}\mathcal{J}_{mn}(t-t_0)dt.
\end{equation}
This allows us to rewrite, at least formally, the Melnikov function as a Fourier series,
\begin{equation}
\label{melfou}
M(t_0)=\sum_{k=-\infty}^\infty e^{2 \pi i k t_0/T}
\int_{-\infty}^\infty \sum_{m \neq n} \beta^1_{mn,k} \frac{d}{ds} \mathcal{J}_{mn}(s) e^{2 \pi i k s/T}ds,
\end{equation}
after the change of the integration variable $s=t-t_0$. Its mean value is thus given by
\begin{equation}
\label{mvalue}
\frac{1}{T}\int_0^T M(t_0)dt_0= \int_{-\infty}^\infty \sum_{m \neq n} \beta^1_{mn,0} \frac{d}{ds}\mathcal{J}_{mn}(s) ds=0.
\end{equation}
In order to check the validity of Fourier series~(\ref{melfou}) we need to bound the norm
\begin{equation}
\label{normfou}
\int_{-\infty}^\infty \left|  \frac{d\mathcal{J}_{mn}}{dt} \right| dt.
\end{equation}
This is not a difficult task because both $p(t)$ and $q(t)$ must be continuously differentiable along the connection and with finite limits when $t \to \pm \infty$, which must coincide with their values at the fixed points. Furthermore, the derivatives $\dot{p}$ and $\dot{q}$ decrease to zero exponentially at infinity due to the hyperbolic character of the fixed points. This implies that the norm~(\ref{normfou}) is necessarily finite.

It is actually simpler to show that the Melnikov function has zero mean. Equation~(\ref{mj}) can be cast into the form
\begin{equation}
M(t_0) = -  \frac{d}{dt_0} \int_{-\infty}^\infty \sum_{m \neq n} \beta^1_{mn}(t+t_0) \mathcal{J}_{mn}(t)dt,
\end{equation}
which, together with the periodicity of the functions $\beta^1_{mn}(t)$, directly provides the desired result. However, the Fourier series setting favors the identification of the conditions under which the Melnikov function vanishes, as Eq.~(\ref{condition}) in the last section, which cannot be straightforwardly extracted from this expression. So we believe that the Fourier series representation will be more helpful in applications to concrete reaction schemes.

In conclusion, we have shown that the Melnikov function of a perturbed instanton connection can always be expanded in Fourier series, provided that it links hyperbolic fixed points. Furthermore, it is always a zero mean function, which implies that it either crosses zero or vanish identically. So for any reaction Hamiltonian we have the following result: given an instanton connection linking two hyperbolic fixed points, any small time-periodic perturbation of the rates will preserve the existence of the connection if the corresponding Melnikov function is not identically zero. In the other case, we would have to study the behavior of the Melnikov distance at higher orders, in order to be able to give a rigorous conclusion~\cite{battelli}.

To finish, let us note the difficulties in extending this result and proving a general theorem about the persistence of the instanton connections even when the Melnikov function is identically zero. In this case, we could recall the expansion of the Melnikov distance in terms of higher-order Melnikov functions~\cite{battelli}. One would be tempted to use an equivalent argument to that of the present section to try to show that an arbitrary-order Melnikov function either has zero mean or vanish identically. This would imply in turn that either some function in the expansion is nonzero, and thus the connection is preserved by means of a transversal crossing of the stable and unstable manifolds, or the whole series becomes identically zero. While this can suggest, at first sight, that the Melnikov distance is also identically zero in this case, we cannot rely on a rigorous argument to prove so, since the perturbative expansion is not analytic in the small parameter in general situations. Furthermore, we can always find a family of transformations (for instance, by continuing the chain of reparametrizations in Eqs.~(\ref{rep1}) and (\ref{rep2})) able to nullify an arbitrary order Melnikov function. So a full proof would need to handle these special perturbations, and this is complicated by the lack of analyticity of the series expansion, despite the probable fact that they are built by combinations of reparametrizations, constant shifts of the parameter values, or other trivial transformations preserving the phase-space topology. Also, proving that the higher-order Melnikov functions have zero mean is not a trivial fact. These imply nonlinear combinations of the external perturbations and the corresponding mixture of Fourier modes that complicates the development of a simple form like Eq.~(\ref{melfou}) in these cases. We illustrate this situation with the second-order Melnikov function in the Appendix. Anyway, perturbations requiring a higher-order Melnikov analysis do not constitute the generic case, and we believe that the majority of the physical situations could be analyzed within the first-order, or at most second-order, formalism.

\section{Discussion}
\label{absence}

So far we have concentrated in showing that the instanton connections persist when the chemical system is temporally forced by a weak perturbation. We devote this section to explain the physical consequences of this fact. As we already argued, the topology of the phase space describes the physics of the system, so if its topology is preserved when the system is forced, this means that the qualitative behavior of the system is preserved too. The disappearance of an instanton connection would mean that this qualitative behavior is modified, but how strongly? These connections are optimal paths linking different physical states (given by the fixed points of the dynamical system), but the absence of one connection communicating two states does not mean that the system cannot go from one to the other. It only means that {\it there is not an optimal way of going}. We will show the importance of this fact using a toy model of biological relevance in plankton modeling. Suppose we have the reactions
\begin{equation}
\label{plnktnrel}
A \stackrel{\gamma}{\longrightarrow}A+A, \qquad A \stackrel{\gamma}{\longrightarrow} \emptyset
\end{equation}
occurring at the same constant rate $\gamma$. This set of equations has been used to model plankton patchiness in some occasions~\cite{zhang,adler,young}. We can straightforwardly write the Schr\"{o}dinger equation for the generating function~\cite{gardiner,baumann},
\begin{equation}
\label{plankton}
\partial_t G = \gamma (p-1)^2 \partial_p G,
\end{equation}
so the reaction Hamiltonian reads
\begin{equation}
H=\gamma (p-1)^2 q.
\end{equation}
The corresponding dynamical system is highly degenerate, with two invariant lines $p=1$ and $q=0$, with all their points being fixed. This means that if we start with an initial condition $(q=n_0,p=1)$, for some $n_0>0$, then this will be the solution for all times. As is totally clear, there are no instanton lines linking the mean-field line with the absorbing-state line. Let us now concentrate directly in the exact PDE~(\ref{plankton}) instead of the approximate dynamical system. This is a linear, first-order PDE which can be easily solved along characteristics~\cite{gardiner}
\begin{equation}
G(p,t)=G_0 \left[1-\frac{1-p}{1+(1-p)\gamma t}\right],
\end{equation}
and due to normalization, we have the long-time asymptotics
\begin{equation}
\lim_{t \to \infty} G(p,t) = \lim_{t \to \infty} G_0 \left[1-\frac{1-p}{1+(1-p)\gamma t}\right] = G_0(1) = 1,
\end{equation}
which corresponds exactly to the probability distribution $P_n=\delta_{n0}$. This means that in the infinite-time limit, regardless of the initial condition, the system will be in the absorbing state with zero particles. So, as we have seen, {\it the absence of instanton connections does not stop the system from undergoing an extinction transition}. One can figure out how this happens representing the process~(\ref{plnktnrel}) with the It\^{o} stochastic differential equation~\cite{adler,escudero1}
\begin{equation}
d\rho=\sqrt{2 \gamma \rho}dW,
\end{equation}
where $W$ denotes a Wiener process. This equation describes Brownian motion with state-dependent diffusion and tells us that the transition to extinction is very different in this case: no optimal trajectory is chosen; instead, the state with zero particles is reached after performing a random walk from the initial condition. This will not necessarily happen in every situation; actually, state-dependent Brownian motion is one of the simplest possibilities. More complicated Hamiltonians with powers of the momentum above the second exhibit properties corresponding to non-Gaussian statistics~\cite{gardiner}. In general, the random walk will be more complex than simple Brownian motion, but all the cases will have in common the absence of an optimal way of going from one physical state to another.

The instanton connections are a useful tool that allows us to calculate the frequency of rare fluctuations that drive the system among different states. The mean transition time is obtained, at exponential order, computing the action along the instanton connection~\cite{elgart}. If our perturbation is pure time reparametrization, as the second perturbation in Eqs.~(\ref{rep1}) and (\ref{rep2}), and the function $\phi(t)$ periodic, a straightforward computation reveals that the action along the heteroclinic orbit is exactly the same as for the unperturbed system. More general perturbations are not tractable analytically, and we would have to rely on a numerical treatment of the problem. The heteroclinic connection can be obtained, for instance, by means of a shooting method and the action computed as a numerical integral on it. In any case, we expect that small, well-behaved, periodic perturbations will yield transition times of the same order of magnitude as the unperturbed system. Instanton persistence is important as it allows approximating this sort of nonautonomous systems by their autonomous counterparts in the first instance and serves as a justification of the quasistationary approximation for slow signals. Perturbations that are not periodic might have a stronger effect on the system, and in particular, singular enough perturbations could change the phase-space topology; however, it is rather difficult to conceive physical situations that give rise to such a perturbation. Also, the mathematical framework changes considerably, since the absence of a Poincar\'{e} map does not allow the definition of the instanton as a connection among fixed points of this map, like in the periodic case. In any case, even if the instanton connections were broken, but the physical states did not drift apart form each other, transitions will be possible if we wait long enough, due to the uncorrelated fluctuations the system is subject to~\cite{lindenberg}, but however, there will be no optimal paths linking those states.

\section{Conclusions and Outlook}

In this work we have studied the persistence of instanton connections in chemical systems, when we promote the reaction rates from constants to functions of time. A set of chemical reactions can be mapped, under the eikonal approximation, onto a dynamical system, the fixed points of which denote the possible states where the chemical system can be found. Connections among fixed points denote optimal transition paths communicating different states. The persistence of these instanton lines indicates the robustsness of the physical processes that are taking place in the reactor and shows us that the qualitative behavior of the system will be similar when subject to small nonautonomous perturbations. We have shown our results with one particular model, and we have stated them in a more general setting when it has been possible. The main theoretical tool that we have employed is the Melnikov function, which has proven itself very useful in dealing with this type of problems.

One of the directions in which we would like to extend the theory is the problematic of connections linking nonhyperbolic, or one nonhyperbolic and one hyperbolic, fixed points. The physical motivation comes from the recent classification of phase transitions in reaction-diffusion models, which identifies critical points of the physical system with the nonhyperbolic fixed points at a bifurcation threshold of the dynamical system~\cite{elgart2}. While it is very difficult to see what happens in the general situation, we know that the mean value of the Melnikov function is zero if the mean value of all the external perturbations is zero, as is shown by Eq.~(\ref{mvalue}), provided it is well defined in this case. This suggests to us that in a broad number of situations the topology of the phase space will remain unchanged if the mean value of all the external perturbations vanishes, even when some of its fixed points were nonhyperbolic. It is, however, very difficult to prove a result in that direction, since perturbating a system in a critical state could have unpredictable consequences. Furthermore, the existence of the Fourier series expansion of the Melnikov function is no longer guaranteed, as the loss of hyperbolicity implies that we do not have an exponential decay at infinity of the absolute value in Eq.~(\ref{normfou}).

If some of the perturbations have a nonzero mean, then the problem becomes much more difficult. The system will be moved out of criticality at some of its points, and its behavior will be more similar to that of the (partially) noncritical phase corresponding to the autonomous system with the parameters retuned according to the mean value of the external perturbations. Unfortunately, we cannot map this situation to that described in Sec.~\ref{ip}, because in this case we have no control on the amplitude of the perturbation, which may be comparable to the distance to some of the critical points. So the problem becomes genuinely nonperturbational, and we can no longer employ a technique like the Melnikov function. In this case it is difficult to deal even with very slow perturbations, since this type of settings favors the appearance of soft modes, which rules out the possibility of treating the external signals adiabatically~\cite{dykman}.

Another problem we would like to deal with is the case of several perturbations with different periods. The simplest case is that in which the ratio of the periods of all possible different pairs of perturbations that are affecting the system is a rational number. In this case, we say that the perturbations are commensurable, and we can find a period which is common to all of them. Once obtained, we have reduced our problem to the one studied in Sec~\ref{ip}. The case of incommensurable perturbations is, of course, more complex, as it implies that the dynamics of the perturbed system cannot be reduced to the dynamics of a periodic Poincar\'{e} map. The fixed points of the unperturbed system give rise to quasiperiodic solutions of the perturbed problem, with their invariant manifolds being quasiperiodic as well~\cite{chow}.

There are many other questions that remain to be answered. One is the possible appearance of chaotic behavior induced by internal stochasticity. As noted in~\cite{assaf}, for two reacting species we have a four-dimensional eikonal Hamiltonian, which will give rise, in general, to a three-dimensional dynamical system on some Riemannian manifold. In this case, chaos, unlike in the two-dimensional mean-field dynamical system, is indeed possible. But chaos is also possible in the case of one reacting species obeying reaction rules with time-dependent rates. In this situation, "energy" is no longer conserved and the nonautonomous dynamical system, without integrals of motion, becomes effectively three dimensional. Indeed, situations like the one plotted in Fig.~(\ref{fase2}), that is, when the heteroclinic connection is preserved due to the intersection of the stable and unstable manifolds, give rise to complex dynamical scenarios. These include the appearance of Smale horseshoes and even the presence of strange attractors related to the creation and destruction of such horseshoes~\cite{smale,mora,pumarino}. Although the generation of chaotic behavior of this type can be attractive from a nonlinear dynamics point of view, we are not aware, at this point, of its possible physical meaning.

Of course, studying multispecies reactions, both autonomous and nonautonomous, is a very interesting problem that deserves further efforts. The nonautonomous situation implies the technical difficulty of extending the Melnikov function theory to a higher dimensionality. The ideas developed for three-dimensional systems~\cite{rodriguez,costal} could perhaps be adapted for studying four- or higher-dimensional problems and to try to obtain in this way the results that we would need to understand the more general multispecies reactions. Also, as we have already pointed out, the Melnikov function is a perturbative result. It allows us to treat arbitrary frequency but necessarily small perturbations. To fully understand the dynamics of nonautonomous chemical reactions, even in the one species case, we would need a nonperturbational result. With it we could try to address questions such as the forcing of systems near criticality or the appearance of soft modes. As we can see, chemical kinetics presents a very complex phenomenology, and it is challenging from a mathematical point of view. A deep understanding of the physics of these nonequilibrium systems will presumably imply the parallel development of powerful methodological techniques.

\section*{Acknowledgments}

The authors gratefully acknowledge input from Ernest Fontich and Carles Sim\'{o}. This work has been partially supported by the Ministerio de Educaci\'{o}n y Ciencia (Spain) through Projects Nos. EX2005-0976, FIS2005-01729, and MTM2005-02094.

\appendix*
\section{Second-order Melnikov function}

Consider the two-dimensional dynamical system
\begin{equation}
\dot{{\bf x}}= {\bf f}({\bf x})+\epsilon {\bf g}({\bf x},t)+\epsilon^2 {\bf h}({\bf x},t),
\end{equation}
where ${\bf x}=(x_1,x_2)$, ${\bf f}=(f_1,f_2)$, ${\bf g}=(g_1,g_2)$, and ${\bf h}=(h_1,h_2)$ are sufficiently regular functions. The second-order Melnikov function reads
\begin{eqnarray}
\nonumber
M_2(t_0)=\int_{-\infty}^\infty {\bf f}({\bf x}_h(t)) \wedge
\left[ 2^{-1} D^2 {\bf f}({\bf x}_h(t))\{ {\bf x}^1_h(t+t_0,t_0) \}^2 \right .\\
\left. +D {\bf g}({\bf x}_h(t),t+t_0) {\bf x}^1_h(t+t_0,t_0) + {\bf h}({\bf x}_h(t),t+t_0) \right]dt,
\label{m2}
\end{eqnarray}
where $D^2 {\bf f}$ and $D {\bf g}$ denote the Hessian and Jacobian of ${\bf f}$ and ${\bf g}$, respectively,
${\bf x}_h(t)$ is the solution parametrizing some given heteroclinic connection, and ${\bf x}^1_h$ is the solution to the variational equation
\begin{equation}
\label{variation}
\dot{{\bf x}}^1_h(t,t_0)= D {\bf f}({\bf x}_h(t-t_0)){\bf x}^1_h(t,t_0)+{\bf g}({\bf x}_h(t-t_0),t).
\end{equation}
We now assume that this dynamical system is a reaction Hamiltonian system and both ${\bf g}$ and ${\bf h}$ come from small nonautonomous perturbations of the Hamiltonian parameters, just like in Sec.~\ref{ip}. Then, because Eq.~(\ref{variation}) is linear, its solution will depend linearly on the Fourier modes of the time-dependent parameters evaluated at $t_0$. But this solution appears quadratically in Eq.~(\ref{m2}) and also multiplying the Jacobian of ${\bf g}$, which depends linearly, by assumption, on the same Fourier modes. This quadratic dependence complicates recasting the second-order Melnikov function into a form similar to Eq.~(\ref{melfou}), which would help enormously for calculating its mean value.

\end{document}